\documentclass[11pt,draftcls,onecolumn]{IEEEtran}
\usepackage{amsmath}
\usepackage{amsfonts}
\usepackage{amssymb}
\usepackage{amsthm}
\usepackage{graphicx}

\newtheorem{Lemma}{Lemma}
\newtheorem{defn}{Definition}
\newtheorem{thm}{Theorem}

\title{A Bayesian Approach to Data Fusion in Sensor Networks}
\author{\IEEEauthorblockN{Zhiyuan Weng, ~~~~~Petar M. Djuri\'{c}*~~~\IEEEmembership{Fellow,~IEEE.}}\\
\IEEEauthorblockA{Department of Electrical and Computer Engineering\\
Stony Brook University, Stony Brook, New York 11794\\
Phone:(631) 632-8423 Fax:  (631) 632-8494\\
Email: zhiyuan.weng@gmail.com, djuric@ece.sunysb.edu\\
EDICS: SEN-FUSE
}}

\begin{document}

%
\maketitle
\begin{abstract}
In this paper, we address the fusion problem in wireless sensor networks, where the cross-correlation between the estimates is unknown. To solve the problem within the Bayesian framework, we assume that the covariance matrix has a prior distribution. We also assume that we know the covariance of each estimate, i.e., the diagonal block of the entire covariance matrix (of the random vector consisting of the two estimates). 
We then derive the conditional distribution of the off-diagonal blocks, which is the cross-correlation of our interest.
We show that when there are two nodes, the conditional distribution happens to be the inverted matrix variate $t$-distribution, from which we can readily sample. For more than two nodes, the conditional distribution is no longer the inverted matrix variate $t$-distribution. But we show that we can decompose it into several sampling problems, each of which is the inverted matrix variate $t$-distribution and therefore we can still sample from it.
Since we can sample from this distribution, it enables us to use the Monte Carlo method to compute the minimum mean square error estimate for the fusion problem.
We use two models to generate experiment data and demonstrate the generality of our method. Simulation results show that the proposed method works better than the popular covariance intersection method.
\end{abstract}
%


\section{Introduction}
In the recent past, research in the area of wireless sensor networks(WSNs) has been steadily growing due to its wide applications. A WSN consists of many sensor nodes that cooperate with each other to perform a measurement or monitoring task, in which data are exchanged and shared between neighbours through wireless communication. The objective of a WSN is to utilize the data at different locations to enhance the measurement performance. With a centralized architecture, a fusion center collects the data from all the sensors to perform the computation and processing task. However, in most cases, a decentralized approach is preferred, because it can provide a degree of scalability and robustness which cannot be achieved with traditional centralized architectures.

Although the notion of decentralization has long been an appeal, exploiting its expected benefits has proven notoriously difficult. In many applications, the information propagated through a sensor network is transformed to a form that provides the estimated state of interest. In many distributed Kalman filter applications \cite{Jinwen2012,cattivelli2010diffusion,olfati2007distributed,olfati2005distributed}, the information is converted into the first and second moment statistics. With the statistics from neighbours at hand, fusing the estimates to obtain a better estimate is expected. A serous problem arising in such setting is the effect of redundant information \cite{hall2001handbook}. The estimates provided by different nodes have unknown cross-correlations. This is particularly true for networks with unknown topological
structure.
Pieces of information from two nodes cannot be simply combined using averaging and weighted averaging unless they are independent or have a known degree of correlation.

Many approaches have been proposed to mitigate the problem. Most of them fall in two categories. The first is looking for an optimal linear combination of estimates in terms of some criterion, for example, weighted least squares or minimum variance \cite{bar1986effect,bar1981track}.
In \cite{li2003optimal, zhu1999best, li2000unified, li2001optimal}, a unified model is developed for estimation fusion based upon the best linear unbiased estimation (BLUE) or linear minimum variance approach.
The second category tries to fuse the available estimates directly \cite{chong1985information, chang1997optimal, drummond1997hybrid, willsky1982combining, miller1998tracklets}. Algorithms for fusing both the first and the second moments for linear systems have been proposed. It is a linear combination of estimates when the first two moments are given.
However, none of the above investigated the situation where the covariance of each estimate is available while the cross covariance are missing.
Consider the following problem.
Given $k$ estimates $x_j$ for $j\in\{1,\cdots,k\}$ of the true state vector $x_0\in\mathbb{R}^{m\times 1}$ with their covariance matrices of the estimation error, $P_{jj}$, we seek a fusion scheme that combines the available information and provides an estimate $\hat{x}_0$ with minimum mean square error. We use $P_0$ to denote the covariance of the estimation error of $\hat{x}_0$.
A naive but simple method is to calculate the weighted average, where the weighting coefficients are proportional to the degrees of the nodes (the numbers of the neighbours of the nodes) \cite{cattivelli2010diffusion}. The approach makes sense because the higher the degree, the more information the node collects and the better it does in estimation.
A more complicated and popular one is known as the covariance intersection method \cite{julier1997non}. It provides a general framework for information fusion with lack of knowledge about cross-correlation between noisy measurements, and it yields consistent estimates between the fused local estimates.
In \cite{julier1997non}, the authors have proposed the covariance intersection method, in which all possible covariance matrix of the resulting estimate is upper bounded by a convex combination of the covariances in the sense that the upper bound matrix minus the covariance matrix of the resulting estimate is a positive semidefinite matrix.
The algorithm can be expressed as
\begin{align*}
P_0^{-1}&=\sum_{j=1}^k\omega_j P_{jj}^{-1}\\
P_0^{-1}\hat{x}_0&=\sum_{j=1}^k\omega_j P_{jj}^{-1}x_j,
\end{align*}
where the weighting coefficient $\omega_j\in[0,1]$ and $\sum_{j=1}^kw_j=1$ hold.
Different performance criteria can be used to decide the value of $\omega_j$. Since the mean square error is of our interests, we use the trace of $P_0$ as the criteria. The minimization of the trace requires iterative minimization of the given nonlinear cost function with respect to the weight coefficients $\omega_j$. In order to reduce the computational complexity, several suboptimal non-iterative algorithms for fast covariance intersection have been developed \cite{niehsen2002information,franken2005improved}.

In \cite{niehsen2002information}, it was reasoned that a replacement of $P_{ii}$ by $P_{jj}$ and vice versa must lead to correspondingly switched coefficients $\omega_i$ and $\omega_j$ and that if $\textrm{tr}(P_{ii})$ $\ll$ $\textrm{tr}(P_{jj})$ for $j\neq i, j\in\{1,\cdots,k\}$ one would expect to get $\omega_i\approx 1$. Thus it was suggested to use the linear equations
\begin{align}
\textrm{tr}(P_{ii})w_i-\textrm{tr}(P_{jj})w_{j}=0, (i,j=1,\cdots,k)
\end{align}
which leads to the solution:
\begin{align}\label{fastci}
\omega_i = \dfrac{1/{\textrm{tr}(P_{ii})}}{\sum_{j=1}^k1/\textrm{tr}(P_{jj})}.
\end{align}
In \cite{franken2005improved}, it was pointed out that the above approximation fails to consider the relative orientation of the estimation error variance matrices which may lead to a degraded performance in certain applications. Accordingly, an improved fast covariance intersection algorithm was proposed which comes with increased computational complexity while yielding better performance in some cases and comparable results in all other ones. (We will use these methods in the sequel for comparison with our algorithm.) In \cite{hurley2002information}, it is pointed out that the covariance intersection is a special case of the generalized fusion, which can be described as
\begin{align}\label{generalizedfusion}
p_0(x)=\dfrac{\prod_{i=1}^k p_i^{\omega_i}(x)}{\int \prod_{i=1}^k p^{\omega_i}_i(x) \textrm{d}x},
\end{align}
where $\sum_{i=1}^k \omega_i=1$; $p_i(x)$ is the distribution at node $i$.

In this work, we use a Bayesian approach to address the problem. In \cite{weng2012}, we investigated the fusing scheme for two nodes. In this paper, we extend the work and deal with the general situation in the wireless sensor network, where information from multiple nodes is to be fused. We assume that the prior of the covariance matrix is the Wishart distribution. Since we know the covariance matrix for each estimate, which is just the diagonal submatrix of the entire covariance matrix, we can derive the conditional distribution of the off-diagonal submatrices.
When there are two nodes, we show that this conditional distribution is the inverted matrix variate $t$-distribution. It is known that one can easily sample from this distribution, entailing that we can efficiently use the Monte Carlo method to compute the minimum mean square error (MMSE) estimate. For multiple nodes, the distribution of off-diagonal blocks are no longer the inverted matrix variate $t$-distribution. But we demonstrate that Bayes' rule can be used to decompose the conditional probability density function (PDF) into a product of several PDFs, each representing an inverted matrix variate $t$-distribution. Therefore, we can still sample from the distribution of off-diagonal blocks in the case of multiple nodes, as well as computing the MMSE estimate.
Our main contribution is solving the problem in the Bayesian framework by using the Monte Carlo methods.
An advantage of our method is that we use matrix weighting coefficients instead of scalar ones, which gives us more freedom to handle the element-wise correlation. Also, we use minimum mean square error as the criterion, which is more popular than the minimax criterion used in the covariance intersection method.
Simulation results show that the proposed method works much better than the traditional covariance intersection method.

The paper is organized as follows. We formulate the problem in Section 2. Sampling methods in the case of two nodes and multiple nodes are discussed in Section 3 and Section 4, respectively. Simulation results of the proposed algorithm are presented in Section 5. Section 6 concludes our paper.

The notation we use in this paper is as follows. Uppercase letters refer to  matrices and lowercase letters to vectors or scalars; $|A|$ is the determinant of a matrix $A$; $A>B$ means that $A-B$ is a positive definite matrix; $x\sim p(x)$ signifies that the random variable $x$ is distributed according to $p(x)$; the symbol $\otimes$ denotes Kronecker product; $I_m$ is the identity matrix with size $m\times m$; $\textrm{tr}(A)$ is the trace of the matrix $A$;  $O$ is a matrix with all entries equal to zero; $\Gamma(\cdot)$ is the standard gamma function, and $\Gamma_l(\cdot)$ is the multivariate gamma function \cite{james1964distributions} defined as
\begin{align}\label{mgamma}
\Gamma_l(n)=\pi^{l(l-1)/4}\prod_{j=1}^l\Gamma\left(n-\frac{1}{2}(j-1)\right).
\end{align}

\section{Problem Formulation}

Consider that a node in a network has $k-1$ nodes in its neighbourhood. By communication with its neighbours, it has $k$ available measurements, including the one from itself. Each measurement $x_j$ for $j\in\{1,\cdots,k\}$ is a $m\times 1$ vector, with the covariance matrices of the estimation error $P_{jj}$. We concatenate the $k$ vectors and let
\begin{align}\label{cov0}
x=\begin{bmatrix}x_1\\
x_2\\
\vdots\\
x_k
\end{bmatrix}
\end{align}
where $x\in\mathbb{R}^{mk\times 1}$.
We assume the mean of $x_j$ is the true state $x_0$. Therefore, the covariance matrix of $x$ is also the covariance matrix of the estimation error of $x$.
We use $P_x$ to denote the covariance matrix of $x$.
\begin{align}\label{px}
P_x=\begin{bmatrix}
P_{11}   & P_{12}   & \cdots & P_{1k}\\
P_{12}^T & P_{22}   & \cdots & P_{2k}\\
\vdots & \vdots     & \ddots & \vdots\\
P_{1k}^T & P_{2k}^T & \cdots & P_{kk}
\end{bmatrix}.
\end{align}
We start by considering linear and unbiased estimator in the form
\begin{align}\label{fmmse}
\hat{x}_0=W^Tx.
\end{align}
$W$ is the weighting coefficient matrix
\begin{align}
W=\begin{bmatrix}W_1^T\\
W_2^T\\
\vdots\\
W_k^T
\end{bmatrix}
\end{align}
where $W_j\in\mathbb{R}^{m\times m}$. Since it should be unbiased, we require
\begin{align}\label{WI_eq_1}
W_1+W_2+\cdots+W_k=I.
\end{align}
Let $I_{(k)}$ be a $km\times m$ matrix concatenated vertically by $k$ identity matrices with size $m\times m$,
\begin{align}
I_{(k)}=\begin{bmatrix}
I_m\\
I_m\\
\vdots\\
I_m
\end{bmatrix}.
\end{align}
Then (\ref{WI_eq_1}) becomes
\begin{align}
W^TI_{(k)}=I.
\end{align}
Let $P_0$ be the covariance matrix of $\hat{x}_0$,
which can be expressed as
\begin{align}
P_0=W^TE(xx^T)W=W^TP_xW.
\end{align}
The minimization of the mean square error is equivalent to the minimization of $\textrm{tr}(P_0)$,
This can be carried out by using the method of Lagrange multipliers. Let
$\Lambda$ be the matrix of Lagrange multipliers. Define now $L$ as
\begin{align}
L(W)=\textrm{tr}(W^TP_xW)+\textrm{tr}(\Lambda(W^TI_k-I))
\end{align}
take derivative with respect to $W$ and $\Lambda$ and using the identity
\begin{align}
\dfrac{\partial ~\textrm{tr}(XAX^T)}{\partial X}&=XA+XA^T\\
\dfrac{\partial ~\textrm{tr}(AXB)}{\partial X}&=A^TB^T,
\end{align}
we obtain the stationary points by the following equations:
\begin{align}
2W^TP_x+\Lambda^TI_{(k)}^T&=0\\
W^TI_{(k)}&=I.
\end{align}
Combining all of the three equations, we obtain
\begin{align}
W^T&=\left(I_{(k)}^TP_x^{-1}I_{(k)}\right)^{-1}I_{(k)}^TP_x^{-1}\label{weight}\\
P_0&=W^TP_xW=\left(I_{(k)}^TP_x^{-1}I_{(k)}\right)^{-1}.
\end{align}
By substituting (\ref{weight}) into (\ref{fmmse}), we have
\begin{align}\label{x0}
\hat{x}_0=\left(I_{(k)}^TP_x^{-1}I_{(k)}\right)^{-1}I_{(k)}^TP_x^{-1}x.
\end{align}

However, in many situations we do not have information about $P_{ij}$ for $i\neq j$. For example, in a sensor network, when two nodes have their measurements and we want to fuse them, we often do not know their cross-covariance.

Our strategy to solving the problem is to put it into a Bayesian framework. We assume that $P_x$ has a prior and that the prior is the Wishart distribution. The Wishart distribution is any of a family of probability distributions defined over symmetric, nonnegative-definite matrix-valued random matrices. These distributions are of great importance in the estimation of covariance matrices in multivariate statistics \cite{anderson2003introduction}.
The Wishart distribution is defined as follows.

The $l\times l$ random matrix $A$ is said to have a Wishart distribution if its probability distribution function (pdf) is given by
\begin{align*}
p(A)=\dfrac{|A|^{\frac{n-l-1}{2}}\exp\left(-\frac{1}{2}\textrm{tr}(\Sigma^{-1}A) \right )}{2^{\frac{kn}{2}}|\Sigma|^{\frac{n}{2}}\Gamma_l(\frac{n}{2})},
\end{align*}
where $\Sigma$ is a positive definite matrix, $n\geq l$ is the degree of freedom and $\Gamma_l$ is defined by \eqref{mgamma}.
We use $\mathcal{W}_l(n,\Sigma)$ to denote the Wishart distribution.
The degree of freedom $n$ also plays an important role in our Bayesian framework as later we will see.
We will omit $l$ and write simply $\mathcal{W}(n,\Sigma)$ if the size of the matrix is obvious from the context.

The Wishart distribution is strongly related to the multivariate normal distribution. Suppose $X$ is an $n\times l$ matrix, the rows of which have $l$-variate normal distribution with zero mean and covariance matrix $\Sigma$, denoted as $\mathcal{N}(0,\Sigma)$. Then the $l\times l$ random matrix $A=X^TX$ has a Wishart distribution, i.e., $\mathcal{W}(n,\Sigma)$. This property makes the generation of Wishart random matrices easy.

We use $P_o$ and $P_d$ to denote the off-diagonal block matrices and the diagonal block matrices, respectively, i.e.,
\begin{align}
P_{d}&=\left\{P_{jj}:~j\in\{1,\cdots,k\}\right\}\\
P_{o}&=\left\{P_{ij}:~i\neq j;~i,j\in\{1,\cdots,k\}\right\}.
\end{align}

In our problem, we know $P_d$. To fuse the data, we would like to have information of $P_o$ conditioned on $P_d$. We express this by the conditional
\begin{align*}
p\left(P_o|P_d\right)&=\dfrac{p(P_x)}{p\left(P_d\right)}.
\end{align*}
Since $P_d$ is known, our weight matrices $W$, and therefore $\hat{x}_0$ are uniquely determined by $P_o$ as in (\ref{x0}). We think of it as a function of the matrix variable $P_o$ and use $f(P_o)$ to denote it. Note that $P_o$ cannot be an arbitrary matrix. $P_o$ must lie in the set $\mathcal{P}_o$ defined by
\begin{align}
\mathcal{P}_o=\left\{P_o:P_x > 0\right\},
\end{align}
where $P_x$ is defined in (\ref{px}).
We express the MMSE estimator by
\begin{align*}
\hat{x}_{mmse}=\int_{\mathcal{P}_o} f(P_o) p(P_o|P_d) d P_o.
\end{align*}
Unfortunately, the above integral is computationally intractable.

In order to approximate the integral, we have to resort to the Monte Carlo method. We sample $M$ independent random matrices, $P^{(j)}_o\sim p(P_o|P_d)$ for $j=1,\cdots,M$. Then the Monte Carlo method approximates $\hat{x}_{mmse}$ by the following expression:
\begin{align*}
\hat{x}_{mmse}\approx\dfrac{1}{M}\sum_{j=1}^M f(P^{(j)}_o).
\end{align*}

An immediate question is how we can sample from the conditional distribution $p(P_o|P_d)$. We answer the question in the next two sections.

\section{Fusion for two nodes}\label{sec2}
In this section, we discuss the sampling method for the conditional distribution of the off-diagonal blocks when there are two nodes. In the case of known $P_o$, the weight matrix $W_1$ and $W_2$ can be expressed as
\begin{align}
W_1 &= (P_{22}-P_{12}^T)(P_{11}-P_{12}-P_{12}^T+P_{22})^{-1}\label{optimal1}\\
W_2 &= (P_{11}-P_{12})(P_{11}-P_{12}-P_{12}^T+P_{22})^{-1}\label{optimal2},
\end{align}
which are the weights for the optimal fusion in the mean square error sense. When we substitute (\ref{optimal1}) and (\ref{optimal2}) back into (\ref{fmmse}), we have
\begin{align}
\hat{x}_0 &= W_1\hat{x}_1+W_2\hat{x}_2.\nonumber\\
&= (P_{22}-P_{12}^T)(P_{11}-P_{12}-P_{12}^T+P_{22})^{-1}\hat{x}_1\nonumber\\
&~~~~~~+(P_{11}-P_{12})(P_{11}-P_{12}-P_{12}^T+P_{22})^{-1}\hat{x}_2.\label{fmmse2}
\end{align}

Suppose that the random matrix $A$ is distributed according to $\mathcal{W}(n,\Sigma)$. Let the partitions of the two positive definite matrices $A$ and $\Sigma$ be denoted by
\begin{align}\label{partition1}
A=\begin{bmatrix}
A_{11} &A_{12} \\
A_{12}^T & A_{22}
\end{bmatrix}~~~~~\Sigma=\begin{bmatrix}
\Sigma_{11} &\Sigma_{12} \\
\Sigma_{12}^T & \Sigma_{22}
\end{bmatrix}.
\end{align}
Here we assume $\Sigma_{12}=O$. Recall that a Wishart matrix variate $A$ can be expressed as $A=X^TX$. $X$ is a Gaussian random matrix, each column of which has the multivariate normal distribution with covariance matrix $\Sigma$. Therefore $\Sigma_{12}=O$ means that the upper part of each column in $X$ is independent of those in the lower part. Our objective is to derive the expression for $p(A_{12}|A_{11},A_{22})$. We will need two properties of the Wishart distribution in our derivation \cite{anderson2003introduction}. To make it general enough, we assume that $A_{11}$ is with size $l_1\times l_1$ and $A_{22}$ is with size $l_2\times l_2$, $l_1+l_2=l$.

\medskip

\begin{Lemma}\label{lem1}
Let $A$ and $\Sigma$ be partitioned into $l_1$ and $l_2$ rows and columns as shown in (\ref{partition1}).
If $A$ is distributed according to $\mathcal{W}_{l_1}(n,\Sigma)$, then $A_{11}$ is distributed according to $\mathcal{W}_{l_2}(n,\Sigma_{11})$.
\end{Lemma}

\begin{Lemma}\label{lem2}
If $\Sigma_{12}=O$ and $A$ is distributed according to $\mathcal{W}(n,\Sigma)$, then $A_{11}$ and $A_{22}$ are independently distributed.
\end{Lemma}

Lemma \ref{lem1} provides the marginal distributions of $p(A_{11})$ and $p(A_{22})$ (they are $\mathcal{W}(n,\Sigma_{11})$ and $\mathcal{W}(n,\Sigma_{22})$, respectively). Lemma \ref{lem2} maintains that $A_{11}$ and $A_{22}$ are independent. Therefore, $p(A_{12}|A_{11},A_{22})$ becomes
\begin{align*}p(A_{12}|A_{11},A_{22})&=\dfrac{p(A)}{p(A_{11},A_{22})}\\
&=\dfrac{p(A)}{p(A_{11})p(A_{22})}.
\end{align*}
With a little algebraic manipulation, we have
\begin{align}
&p(A_{12}|A_{11},A_{22})\nonumber\\
=&Z\cdot|A|^{\frac{n-l-1}{2}}\nonumber\\
=&Z\cdot\left(|A_{11}||A_{22}-A_{12}^TA_{11}^{-1}A_{12}|\right)^{\frac{n-l-1}{2}}\nonumber\\
=&Z\cdot\left(|A_{11}A_{22}||I-A_{22}^{-1}A_{12}^TA_{11}^{-1}A_{12}|\right)^{\frac{n-l-1}{2}}\label{conpdf1},
\end{align}
where $n>l-1$, and the constant $Z$ equals
\begin{align}
Z&=\dfrac{\left(\prod_{i=1}^{l_1}\Gamma(\frac{1}{2}(n+1-i))\prod_{h=1}^{l_2}\Gamma(\frac{1}{2}(n+1-h))\right)}{\prod_{j=1}^{l}\Gamma(\frac{1}{2}(n+1-j))}\cdot\dfrac{1}{\pi^\frac{l_1l_2}{2}|A_{11}|^{\frac{n-l_1-1}{2}}|A_{22}|^{\frac{n-l_2-1}{2}}}\\
&=\dfrac{\prod_{i=1}^{l_2}\Gamma(\frac{1}{2}(n+1-i))}{\prod_{j=1+l_1}^{l}\Gamma(\frac{1}{2}(n+1-j))}\cdot\dfrac{1}{\pi^\frac{l_1l_2}{2}|A_{11}|^{\frac{n-l_1-1}{2}}|A_{22}|^{\frac{n-l_2-1}{2}}}\\
&=\dfrac{\prod_{i=1}^{l_2}\Gamma(\frac{1}{2}(n+1-i))}{\prod_{j=1}^{l-l_1}\Gamma(\frac{1}{2}(n-l_1+1-j))}\cdot\dfrac{1}{\pi^\frac{l_1l_2}{2}|A_{11}|^{\frac{n-l_1-1}{2}}|A_{22}|^{\frac{n-l_2-1}{2}}}\\
&=\dfrac{\Gamma_{l_2}(\frac{n}{2})}{\Gamma_{l_2}(\frac{1}{2}(n-l_1))}\cdot\dfrac{1}{\pi^\frac{l_1l_2}{2}|A_{11}|^{\frac{n-l_1-1}{2}}|A_{22}|^{\frac{n-l_2-1}{2}}}.
\end{align}
The above distribution is the inverted matrix variate $t$-distribution whose definition is as follows \cite{gupta2000matrix}:

\begin{defn}
The random matrix $T\in\mathbb{R}^{l\times m}$ is said to have an inverted matrix variate $t$-distribution with parameters $M\in\mathbb{R}^{l\times m}$, $\Sigma\in\mathbb{R}^{l\times l}$, $\Omega\in\mathbb{R}^{m\times m}$ and $n$ if its pdf is given by
\begin{align*}
p(T)&=\dfrac{\Gamma_l(\frac{1}{2}(n+m+l-1))}{\pi^{\frac{ml}{2}}\Gamma_l(\frac{1}{2}(n+l-1))}|\Sigma|^{-\frac{m}{2}}|\Omega|^{-\frac{l}{2}}\\
&~~~~~~~~~~~~~|I-\Sigma^{-1}(T-M)\Omega^{-1}(T-M)^T|^{\frac{n-2}{2}},
\end{align*}
where $\Omega>0$, $\Sigma>0$, $n>0$ and $I-\Sigma^{-1}(T-M)\Omega^{-1}(T-M)^T>0$. We denote this by $T\sim\mathcal{IT}_{l,m}(n,M,\Sigma,\Omega)$.
\end{defn}

For our case in (\ref{conpdf1}), it is not difficult to obtain that
\begin{align}\label{conpdf2}
A_{12}^T|A_{11},A_{22}~\sim~\mathcal{IT}_{l_2,l_1}(n-l+1,O,A_{22},A_{11}).
\end{align}

For sampling from the inverted matrix variate $t$-distribution, we use the following lemma \cite{gupta2000matrix}:
\begin{Lemma}\label{lem3}
Let $S\sim\mathcal{W}_l(n+l-1,I_l)$ and $X\sim\mathcal{N}_{l,m}(0,I_l\otimes I_m)$ be independently distributed. For $M\in\mathbb{R}^{l\times m}$, define
\begin{align*}
T = \Sigma^{\frac{1}{2}}(S+XX^T)^{-\frac{1}{2}}X\Omega^{\frac{1}{2}}+M,
\end{align*}
where $S+XX^T=(S+XX^T)^{\frac{1}{2}}((S+XX^T)^{\frac{1}{2}})^T$ and $\Sigma^{\frac{1}{2}}$ and $\Omega^{\frac{1}{2}}$ are the symmetric square roots of the positive definite matrices $\Sigma$ and $\Omega$, respectively. Then, $T\sim\mathcal{IT}_{l,m}(n,M,\Sigma,\Omega)$.
\end{Lemma}

According to Lemma \ref{lem3}, the following theorem follows immediately.
\begin{thm}\label{thm1}
Let random matrices $S\sim\mathcal{W}_{l_2}(n-l_1,I_{l_2})$ and $X\sim\mathcal{N}_{l_2,l_1}(0,I_{l_2}\otimes I_{l_1})$. If
\begin{align*}
A_{12}^T = (A_{22})^{\frac{1}{2}}(S+XX^T)^{-\frac{1}{2}}X(A_{11})^{\frac{1}{2}},
\end{align*}
then $A_{12}^T\sim p\left(A_{12}|A_{11},A_{22}\right)$.
\end{thm}

\textbf{Remark.} We can see that the hyperparameter $\Sigma$ in the prior disappears in the condition distribution as long as it is a block diagonal matrix. On the other hand, the degree of freedom $n$ reflects the prior belief on correlation between the two estimates. This can be used to exploit available information to allow for better estimation.

\section{Fusion for more than two nodes}
In this section, we consider the situation when we have three or more nodes.
For multiple nodes, the conditional distribution of the off-diagonal submatrices is not inverted matrix variate t-distribution, and there is no way to directly sample from it. However we can do it as follows.

Suppose we have $k$ nodes, each measurement is a $m\times 1$ vector. The covariance matrix is
\begin{align}
A=\begin{bmatrix}
A_{11} & A_{12} &  \cdots & A_{1k}\\
A^T_{12} & A_{22} & \cdots & A_{2k}\\
\vdots & \vdots & \ddots & \vdots\\
A^T_{1k} & A^T_{2k} &  \cdots & A_{kk}
\end{bmatrix},
\end{align}
where $A_{jj}\in\mathbb{R}^{m\times m}$. We use $B_j$ to denote
\begin{align}
B_j&=\begin{bmatrix}
A_{11}   & A_{12}   & \cdots  & A_{1j}\\
A^T_{12} & A_{22}   & \cdots  & A_{2j}\\
\vdots   & \vdots   & \ddots  & \vdots\\
A^T_{1j} & A^T_{2j} & \cdots  & A_{jj}
\end{bmatrix}.
\end{align}

The conditional distribution becomes
\begin{align}
&p\left(A_{12},A_{13},A_{23},\cdots,A_{(k-1)k}\right|A_{11},\cdots,A_{kk})\\
=&\dfrac{p(A)}{p(A_{11})p(A_{22})\cdots p(A_{kk})}
\end{align}
which there is no existing method to sample from.
By repeatedly invoking Bayes chain rule, we can write it in this way.
\begin{align}
&p\left(A_{12},A_{13},A_{23},\cdots,A_{(k-1)k}\right|A_{11},\cdots,A_{kk})\\
=&p\left(A_{12}|A_{11},\cdots,A_{kk}\right)p\left(A_{13},A_{23},\cdots,A_{(k-1)k}|A_{12},A_{11},\cdots,A_{kk}\right)\\
=&p\left(A_{12}|A_{11},\cdots,A_{kk}\right)p\left(A_{13},A_{23}|A_{12},A_{11},\cdots,A_{kk}\right)\nonumber\\
&~~~~~~~~~~p\left(A_{14},A_{24},A_{34},\cdots,A_{(k-1)k}|A_{13},A_{23},A_{12},A_{11},\cdots,A_{kk}\right)\\
=&p\left(A_{12}|A_{11},\cdots,A_{kk}\right)p\left(A_{13},A_{23}|A_{12},A_{11},\cdots,A_{kk}\right)\nonumber\\
&~~~~~~~~~~p\left(A_{14},A_{24},A_{34},\cdots,A_{(k-1)k}|B_{3},A_{44},\cdots,A_{kk}\right)\\
=&p\left(A_{12}|A_{11},\cdots,A_{kk}\right)\nonumber\\
&p\left(A_{13},A_{23}|B_{2},A_{33},\cdots,A_{kk}\right)\nonumber\\
&\cdots\nonumber\\
&p\left(A_{1j},A_{2j},\cdots,A_{(j-1)j}|B_{j-1},A_{jj},\cdots,A_{kk}\right)\nonumber\\
&\cdots\nonumber\\
&p\left(A_{1k},\cdots,A_{(k-1)k}|B_{k-1},A_{kk}\right)\label{cascade_form}
\end{align}

Note that according to Lemma \ref{lem1}, we can simplify the conditional distribution that
\begin{align}
p\left(A_{12}|A_{11},\cdots,A_{kk}\right)=p\left(A_{12}|A_{11},A_{22}\right)
\end{align}
and in general
\begin{align}
&p\left(A_{1j},A_{2j},\cdots,A_{(j-1)j}|B_{j-1},A_{jj},\cdots,A_{kk}\right)\\
=&p\left(A_{1j},A_{2j},\cdots,A_{(j-1)j}|B_{j-1},A_{jj}\right).
\end{align}

Therefore (\ref{cascade_form}) becomes
\begin{align}
&p\left(A_{12}|A_{11},A_{22}\right)\label{s1}\\
&p\left(A_{13},A_{23}|B_{2},A_{33}\right)\label{s2}\\
&\cdots\nonumber\\
&p\left(A_{1j},A_{2j},\cdots,A_{(j-1)j}|B_{j-1},A_{jj}\right)\label{s3}\\
&\cdots\nonumber\\
&p\left(A_{1k},\cdots,A_{(k-1)k}|B_{k-1},A_{kk}\right)\label{samp_form}.
\end{align}
Now things becomes easy for us since each factor in (\ref{s1}-\ref{samp_form}) is the inverted matrix variate $t$-distribution, which can be easily sample from. Specifically, we can do it as follows.

According to Theorem \ref{thm1}, we can easily sample $A_{12}$ according to (\ref{s1}). Then we sample $A_{13},A_{23}$ from (\ref{s2}). Let random matrices $S\sim\mathcal{W}_{m}(n-2m,I_{m})$ and $X\sim\mathcal{N}_{m,2m}(0,I_{m}\otimes I_{2m})$. Let
\begin{align}
\begin{bmatrix}A_{13}^T\\ A_{23}^T\end{bmatrix} = A_{33}^{\frac{1}{2}}(S+XX^T)^{-\frac{1}{2}}XB_{2}^{\frac{1}{2}},
\end{align}
where
\begin{align}
B_2&=\begin{bmatrix}
A_{11} & A_{12}\\
A^T_{12} & A_{22}
\end{bmatrix}.
\end{align}
then $[A_{13}, A_{23}]\sim p\left(A_{13},A_{23}|B_{2},A_{33}\right)$.

The iteration goes on for $k-1$ times.
In $j$th iteration,
we sample $A_{1(j+1)},A_{2(j+1)},\cdots, A_{j(j+1)}$ from (\ref{s3}). Let random matrices $S\sim\mathcal{W}_{m}(n-jm,I_{m})$ and $X\sim\mathcal{N}_{m,jm}(0,I_{m}\otimes I_{jm})$. Let
\begin{align}
\begin{bmatrix}
A_{1(j+1)}^T\\
A_{2(j+1)}^T\\
\vdots\\
A_{j(j+1)}^T
\end{bmatrix} = A_{(j+1)(j+1)}^{\frac{1}{2}}(S+XX^T)^{-\frac{1}{2}}XB_{j}^{\frac{1}{2}},
\end{align}
then
\begin{align}
\begin{bmatrix}
A_{1(j+1)}\\
A_{2(j+1)}\\
\vdots\\
A_{j(j+1)}
\end{bmatrix}\sim p\left(A_{1(j+1)},A_{2(j+1)},\cdots,A_{j(j+1)}|B_j,A_{(j+1)(j+1)}\right).
\end{align}

\begin{figure}[h]
\centering
\includegraphics[scale=1, trim=0 50 50 50]{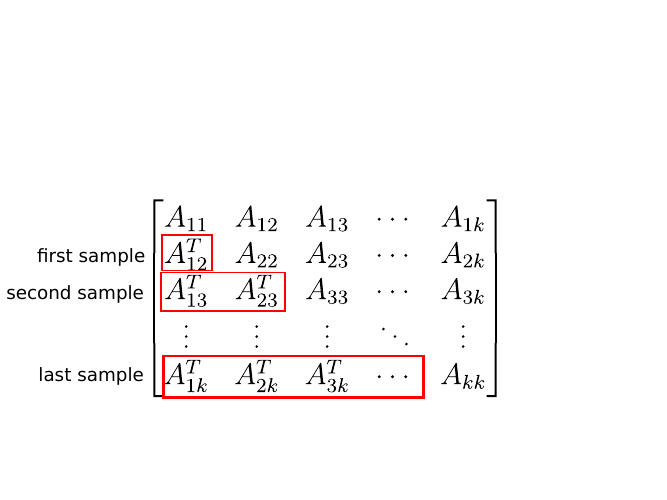}\\
\caption{Illustration of the order of sampling the off-diagonal block matrices.}\label{samplefig}
\end{figure}

Figure \ref{samplefig} shows the steps of the sampling algorithm.
Before ending this section, we wish to emphasize that in multiple nodes situation, fusing two nodes at a time using the method discussed in Section \ref{sec2} will not work. By 'not work', we mean that fusing two nodes at a time and repeatedly do this for multiple nodes is not equivalent to fusing multiple nodes at a time.

\section{Numerical Experiments}

In this section, we construct two models to test our algorithm. The first model is a random matrix with a Wishart distribution (referred to as model 1). The second model is borrowed from a setting which arises in the distributed Kalman filter (referred to as model 2).

Suppose the true state $x_0$ is a zero vector with $2$ elements. We have $k$ available measurements $x_i$ for $i\in\{1,\cdots,k\}$. The measurements have normal distribution with covariance matrix $P_x$. We generate $P_x$ according to $\mathcal{W}(n,\Sigma)$ in each run, where $n=3k$. Since $x_0$ is assumed to be zero, the measurements are with zero mean. We carry out the experiment as follows. For each run, we first generate $P_x$ according to the Wishart distribution and then sample from the corresponding normal distribution to get sample of $x_i$. We suppose the diagonal blocks of $P_x$ are known. Then the proposed method is used to calculate the weighting coefficients and $x_j$'s are merged. We use $100$ samples to estimate the integral (M=100). Finally, we compare $\hat{x}_0$ with $x_0$, which is zero, to measure the performance. We ran the simulation $10000$ times to get the averaged performance. 

\begin{figure}[h]
\centering
\includegraphics[scale=0.9, trim=190 260 190 260]{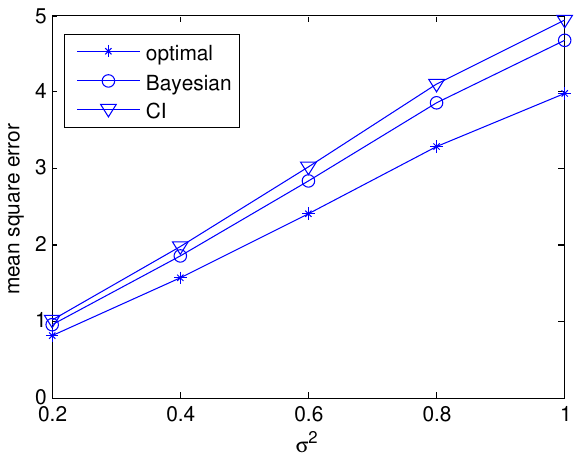}
\caption{The mean square error of the three estimators for different $\sigma^2$ in two nodes situation. (model 1)}\label{m1_1}
\end{figure}

\begin{figure}[h]
\centering
\includegraphics[scale=0.9, trim=190 260 190 260]{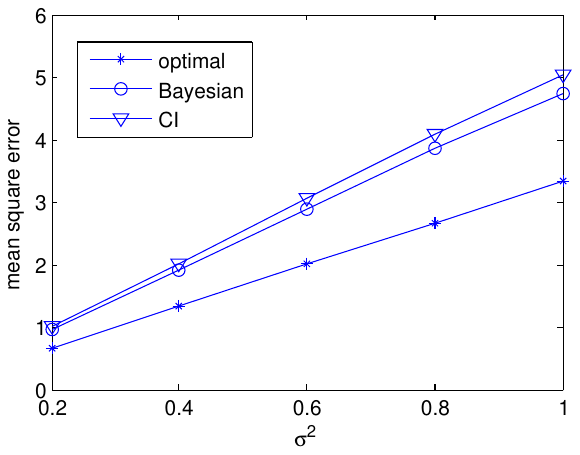}
\caption{The mean square error of the three estimators for different $\sigma^2$ in three nodes situation. (model 1)}\label{m1_2}
\end{figure}

\begin{figure}[h]
\centering
\includegraphics[scale=0.9, trim=190 260 190 260]{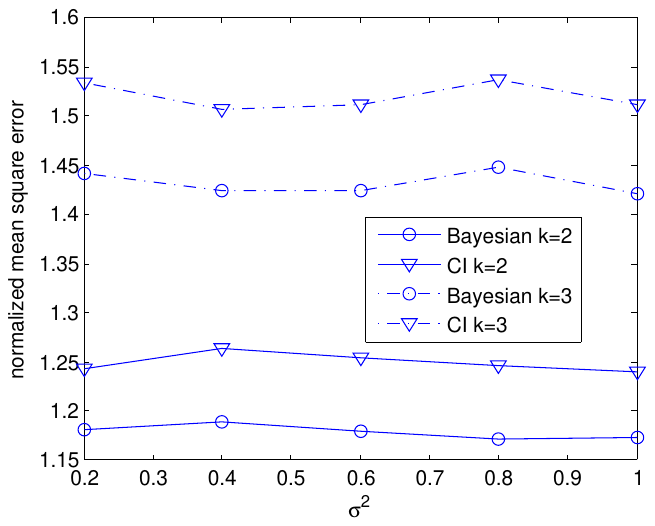}
\caption{The normalized mean square error of the two estimators for different $\sigma^2$ in two and three nodes situation. (model 1)}\label{m1_norm}
\end{figure}

Figure \ref{m1_1} shows the mean square error performance for two nodes and Fig. \ref{m1_2} shows the performance for three nodes. The proposed method is roughly $10\%$ better than the covariance intersection method in both situations. Fig. \ref{m1_norm} shows the normalized mean square error performance, which is obtained by normalizing the MSE of the estimator, using the MSE of the optimal estimator as a measure of scale. We see that in both situations, the proposed estimator outperforms the other. However with the number of nodes growing, the gap between the optimal estimator and the others becomes larger.

Next, we test the case for the second model, the typical distributed Kalman filter case.
Suppose the variable to be estimated is $x_0$ and it has distribution $\mathcal{N}(\mu_0,\Sigma_0)$. The measurements $x_i$ has the conditional distributions $\mathcal{N}(x_0,\Sigma_i)$ for $i\in\{1,\cdots,k\}$. The noise is usually assumed to be independent of each other. We can consider $x_i$ to be measurements as well as estimates since we shall let $\hat{x}_i=x_i$ if we make estimation only based on $x_i$. If we concatenate $k$ measurements into one vector, the distribution of the vector conditioned on $x_0$ is

\begin{align}
\left.\begin{bmatrix}
x_1\\
x_2\\
\vdots\\
x_k
\end{bmatrix}\right|x_0\sim\mathcal{N}\left(\begin{bmatrix}
x_0\\
x_0\\
\vdots\\
x_0
\end{bmatrix},\begin{bmatrix}
\Sigma_1 & O & \cdots & O\\
O&\Sigma_2&\cdots&O\\
\vdots & \vdots & \ddots & \vdots\\
O & O & \cdots & \Sigma_k
\end{bmatrix}
\right).
\end{align}
Furthermore, we can easily obtain its marginal distribution, or
\begin{align}\label{cov11}
\begin{bmatrix}
x_1\\
x_2\\
\vdots\\
x_k
\end{bmatrix}\sim\mathcal{N}\left(\begin{bmatrix}
\mu_0\\
\mu_0\\
\vdots\\
\mu_0
\end{bmatrix},\begin{bmatrix}
\Sigma_0+\Sigma_1 & \Sigma_0 & \cdots & \Sigma_0\\
\Sigma_0 & \Sigma_0+\Sigma_2 & \cdots & \Sigma_0\\
\vdots   & \vdots            & \ddots & \vdots\\
\Sigma_0 & \Sigma_0          & \cdots & \Sigma_0+\Sigma_k
\end{bmatrix}
\right).
\end{align}

Note that the covariance matrix in (\ref{cov11}) is just the one in (\ref{cov0}), which is of our interest. So $P_{ii}=\Sigma_i+\Sigma_0$, and they are known exactly. On the other hand, $P_{ij}=\Sigma_0$ for $i\neq j$ is unknown as well as $\Sigma_i$ for $i\in\{1,\cdots,k\}$.

To generate the data for our numerical experiment, we first draw $\Sigma_0$ from $\mathcal{W}_{2}(3,\sigma_0^2 I_{2})$ and $\Sigma_1$, $\cdots$, $\Sigma_k$ from $\mathcal{W}_{2}(3,\sigma^2 I_{2})$. Then we generate the true value $x_0$ by sampling from $\mathcal{N}(0,\Sigma_0)$. Similarly we generate the measurements $x_1$ and $x_2$ from $\mathcal{N}(x_0,\Sigma_1)$ and $\mathcal{N}(x_0,\Sigma_2)$, respectively. As stated above, the marginal covariance matrix (\ref{cov11}) of the combined measurements becomes
\begin{align}P_x=
\begin{bmatrix}
\Sigma_0+\Sigma_1 & \Sigma_0 & \cdots & \Sigma_0\\
\Sigma_0 & \Sigma_0+\Sigma_2 & \cdots & \Sigma_0\\
\vdots   & \vdots            & \ddots & \vdots\\
\Sigma_0 & \Sigma_0          & \cdots & \Sigma_0+\Sigma_k
\end{bmatrix}.\label{cov2}
\end{align}

Now we have all the data we need for testing and comparing the estimators. For comparison, we use two other estimators, the optimal estimator (\ref{fmmse2}) with all the available  information and the fast covariance intersection method from \cite{franken2005improved}. For each configuration, we ran $10000$ tests. In the proposed algorithm, for calculating $\hat{x}_{mmse}$ we generated $100$ samples. In the legend, we use \emph{optimal, Bayesian} and \emph{CI} to indicate the optimal method, the proposed method, and the fast covariance intersection method, respectively. We would like to point out that the covariance matrix $P_x$ in the simulation does not have the Wishart distribution, as the off-diagonal block matrix is always symmetric. Nevertheless, we will see our estimator still performs well.

\begin{figure}[h]
\centering
\includegraphics[scale=0.9, trim=190 260 190 260]{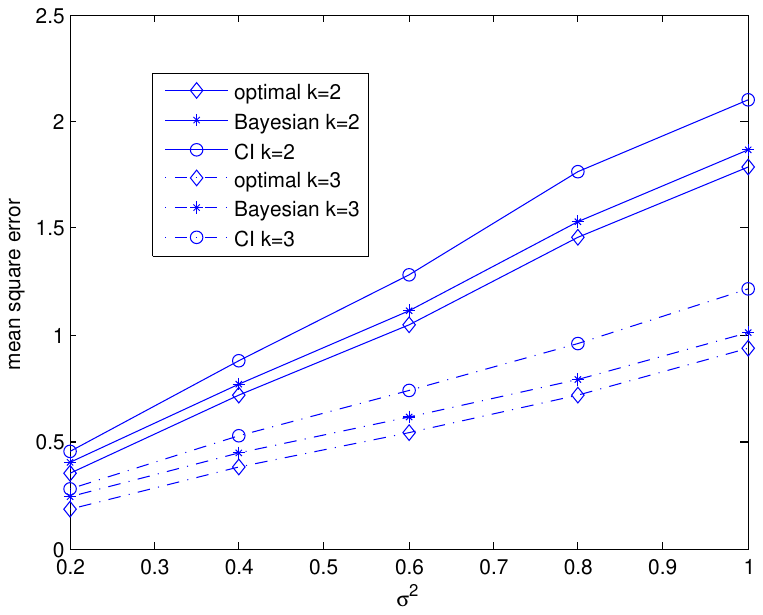}\\
\caption{The mean square error of the three estimators for different $\sigma_0^2$ in two and three nodes situation. ($\sigma^2=1$, model 2)}\label{fig1}
\end{figure}
\begin{figure}[h]
\centering
\includegraphics[scale=0.9, trim=190 260 190 260]{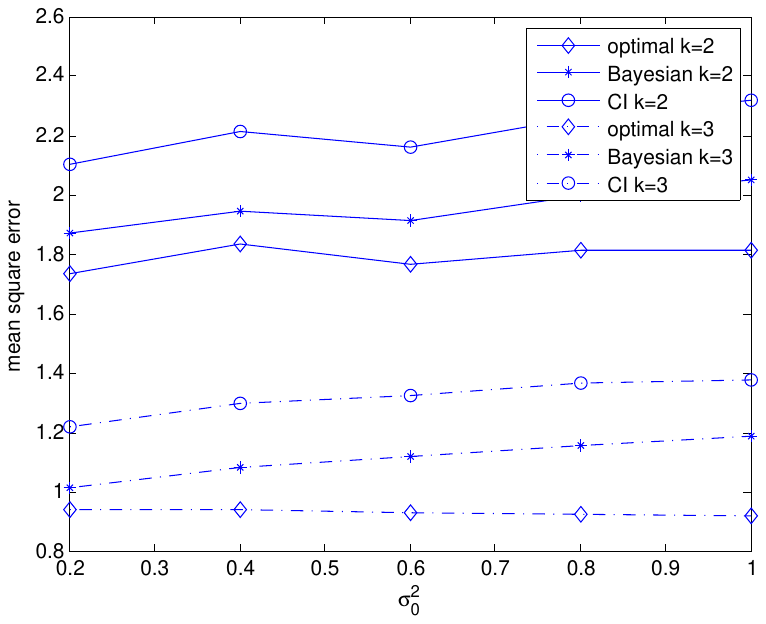}\\
\caption{The mean square error of the three estimators for different $\sigma^2$ in two and three nodes situation. ($\sigma^2_0=0.2$, model 2)}\label{fig2}
\end{figure}

Fig.~\ref{fig1} shows the mean square error of the three estimators for different $\sigma^2_0$. From (\ref{cov2}) we can see that the cross-covariance is determined by $\sigma^2_0$. Roughly speaking, the `larger' the matrix values are, the `more' the estimates relate to each other. From Fig.~\ref{fig1}, we see that the optimal method works best as expected. The proposed Bayesian algorithm is about $20$ percent worse than the optimal one, but much better than the covariance intersection estimator. Fig.~\ref{fig2} shows the mean square error versus different values of $\sigma^2$. Unlike $\sigma^2_0$, $\sigma^2$ has no effect on the cross-covariance. We have similar performance as in the first figure. Again, the optimal estimator is the best, and the proposed estimator has performance that is close to that of the optimal estimator and much better than the performance of the covariance intersection method. From Fig.~\ref{fig1} and Fig.~\ref{fig2}, we can also notice that three nodes lead to better estimation than two nodes. This is contrary to that of model 1. The reason is that in model 1, the number of elements in the off-diagonal blocks increases, which increases the number of dimensions of the variable space. On the other hand, in model 2, the off-diagonal blocks are all $\Sigma_0$. As a result, the number of dimensions does not increase.

\begin{figure}[h]
\centering
\includegraphics[scale=0.9, trim=190 260 190 260]{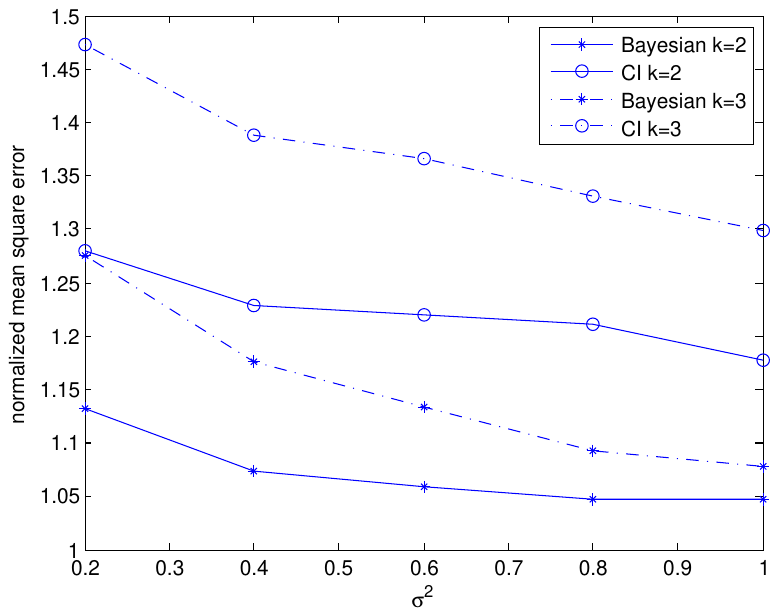}\\
\caption{The normalized mean square error of the three estimators for different $\sigma^2$ in two and three nodes situation. ($\sigma^2_0=0.2$, model 2)}\label{m2_norm_sig}
\end{figure}
\begin{figure}[h]
\centering
\includegraphics[scale=0.9, trim=190 260 190 260]{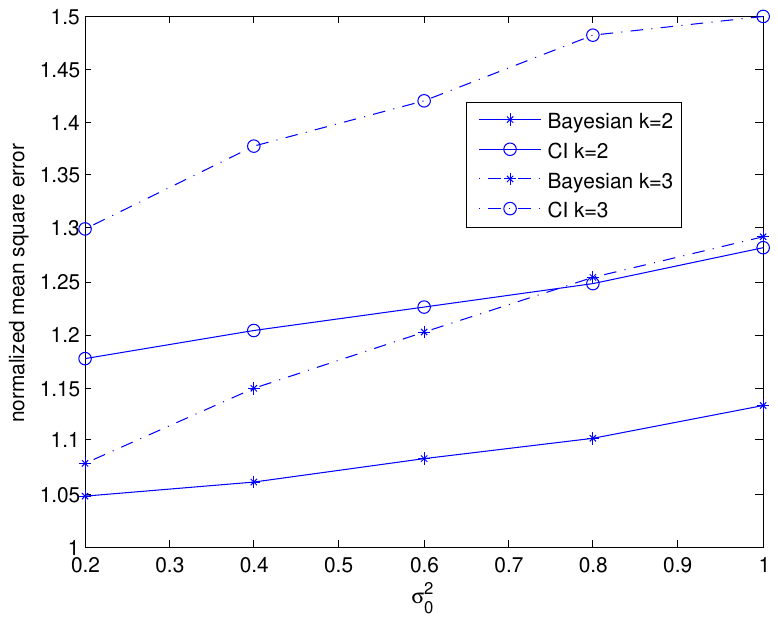}\\
\caption{The normalized mean square error of the three estimators for different $\sigma_0^2$ in two and three nodes situation. ($\sigma^2_0=0.2$, model 2)}\label{m2_norm_sig0}
\end{figure}

\begin{figure}[h]
\centering
\includegraphics[scale=0.9, trim=190 260 190 260]{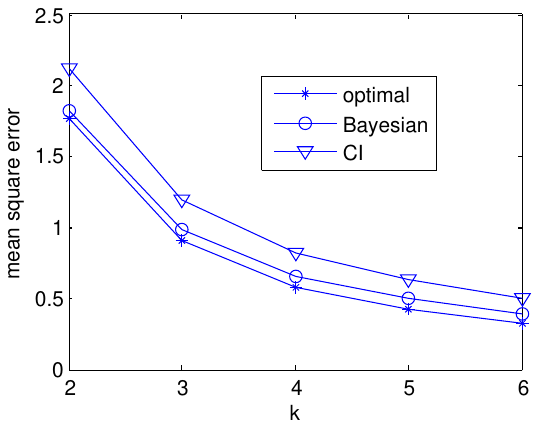}\\
\caption{The mean square error of the two estimators for different number of nodes, ($\sigma^2=1$ for model 1, $\sigma^2=1$, $\sigma^2_0=0.2$ for model 2)}\label{m12_norm_n26}
\end{figure}

\begin{figure}[h]
\centering
\includegraphics[scale=0.9, trim=190 260 190 260]{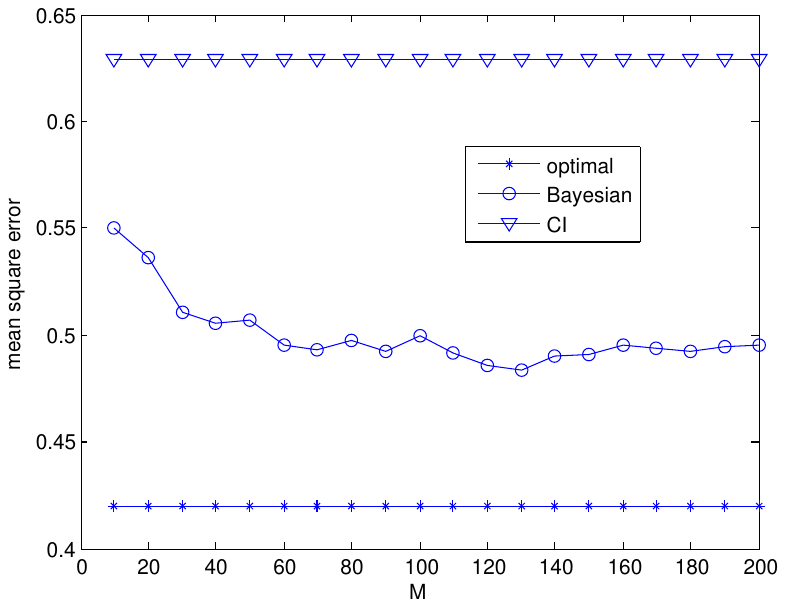}\\
\caption{The mean square error of the estimators for different number of $M$, ($\sigma^2=1$, $\sigma^2_0=0.2$, model 2)}\label{Mtest}
\end{figure}

Fig.~\ref{m2_norm_sig} and Fig.~\ref{m2_norm_sig0} shows the normalized mean square error of the three estimators for different $\sigma^2$. We can see the normalized performance deteriorates with the increase in the number of nodes. Also, the normalized performance becomes better with the increase of $\sigma^2$ but worse with the increase of $\sigma_0^2$.
In Fig. \ref{m12_norm_n26}, the mean square error performance with different $k$ is illustrated for both models. Apparently, the proposed algorithm works better in model 2.
In Fig. \ref{Mtest}, we show the mean square error performance of the Bayesian estimator for different number of sample size, $M$, in the situation of five nodes.
we can see that $100$ samples are almost enough for the Monte Carlo method to obtain the accurate integral estimation. Therefore, the computational complexity of our method is acceptable.


\section{Discussion}

In this article, we propose a Bayesian approach to solve the data fusion problem in wireless sensor network when the cross-covariance between the estimates is not available. We first assume that the prior of the covariance matrix is the Wishart distribution. Because we know the covariance of each estimate, which is the diagonal block of the covariance matrix, we can obtain the conditional distribution of the off-diagonal block. For the case of two nodes, the conditional distribution of this block is the inverted matrix variate $t$-distribution. We also show how to sample from this  distribution. For the case of multiple nodes, the conditional distribution becomes much more complicated and there is no direct way to sample from it. We use Bayes' chain rule to decompose the distribution into a product of several inverted matrix variate $t$-distribution so that we can still sample from it. As a result, we can use the Monte Carlo method to compute the MMSE estimator. Numerical experiments show that the performance of our method is better than that of the covariance intersection method. Another advantage of our algorithm is that under the Bayesian framework, we can modify the hyperparameter of the prior, the degree of freedom $n$, according to the available prior information, to make the algorithm perform better in some special cases.

The curious reader may wonder why we assume the parameter $\Sigma$ of the prior Wishart distribution $\mathcal{W}(n,\Sigma)$ to be a block diagonal matrix. The reason is that by doing so, the diagonal blocks of the resulting covariance matrix are independent from each other. Otherwise, the joint distribution of the diagonal blocks are very complicated making the derivation of the conditional distribution of the off-diagonal blocks very difficult, if not impossible. We can see in the numerical experiment that the Wishart distribution with block diagonal parameter matrix $\Sigma$ is still general enough to allow for good performance. However if we can extend $\Sigma$ to general positive definite matrix, it would give us more freedom to manipulate the prior according available information. This should be the direction of the future efforts.

We wish to emphasize that we never mean to say that the proposed method is more advantageous than the covariance intersection method.
In fact, they are quite different.
Because the covariance intersection uses the minimax criterion and our method uses the minimum mean square error criterion, it makes no sense to say either one is better.
Furthermore, the covariance intersection is a special case of the generalized fusion \eqref{generalizedfusion}. This makes it suitable for multiple fusions without performance deterioration. In other words, the result stays the same if we carry out the fusion multiple times. This property is particularly useful for consensus in networks. On the other hand, the proposed method is not in the framework of the generalized fusion. Multiple fusions by using the proposed method leads to overconfident estimation, i.e., the covariance matrix shrinks each time the operation is carried out. The purpose of our method is to provide an alternative to dealing with the difficult fusion issue in wireless sensor networks.


\bibliographystyle{IEEEbib}
\bibliography{refs}

\begin{thebibliography}{10}

\bibitem{Jinwen2012}
J~Hu, L~Xie, and C~Zhang,
\newblock ``Diffusion {Kalman} filtering based on covariance intersection,''
\newblock {\em Signal Processing, IEEE Transactions on}, vol. 60, no. 2, pp.
  891 --902, feb. 2012.

\bibitem{cattivelli2010diffusion}
F.S. Cattivelli and A.H. Sayed,
\newblock ``Diffusion strategies for distributed {Kalman} filtering and
  smoothing,''
\newblock {\em Automatic Control, IEEE Transactions on}, vol. 55, no. 9, pp.
  2069--2084, 2010.

\bibitem{olfati2007distributed}
R.~Olfati-Saber,
\newblock ``Distributed {Kalman} filtering for sensor networks,''
\newblock in {\em Decision and Control, 2007 46th IEEE Conference on}. IEEE,
  2007, pp. 5492--5498.

\bibitem{olfati2005distributed}
R.~Olfati-Saber,
\newblock ``Distributed {Kalman} filter with embedded consensus filters,''
\newblock in {\em Decision and Control, 2005 and 2005 European Control
  Conference. CDC-ECC'05. 44th IEEE Conference on}. IEEE, 2005, pp. 8179--8184.

\bibitem{hall2001handbook}
D.L. Hall and J.~Llinas,
\newblock {\em Handbook of multisensor data fusion},
\newblock CRC Pr I Llc, 2001.

\bibitem{bar1986effect}
Y.~Bar-Shalom and L.~Campo,
\newblock ``The effect of the common process noise on the two-sensor
  fused-track covariance,''
\newblock {\em Aerospace and Electronic Systems, IEEE Transactions on}, , no.
  6, pp. 803--805, 1986.

\bibitem{bar1981track}
Y.~Bar-Shalom,
\newblock ``On the track-to-track correlation problem,''
\newblock {\em Automatic Control, IEEE Transactions on}, vol. 26, no. 2, pp.
  571--572, 1981.

\bibitem{li2003optimal}
X.R. Li, Y.~Zhu, J.~Wang, and C.~Han,
\newblock ``Optimal linear estimation fusion. i. unified fusion rules,''
\newblock {\em Information Theory, IEEE Transactions on}, vol. 49, no. 9, pp.
  2192--2208, 2003.

\bibitem{zhu1999best}
Y.M. Zhu and X.R. Li,
\newblock ``Best linear unbiased estimation fusion,''
\newblock in {\em Proc. 1999 International Conf. on Information Fusion}, 1999,
  pp. 1054--1061.

\bibitem{li2000unified}
X.R. Li and J.~Wang,
\newblock ``Unified optimal linear estimation fusion¡ªpart ii: Discussions and
  examples,''
\newblock in {\em Proc. 2000 International Conf. on Information Fusion}, 2000.

\bibitem{li2001optimal}
X.R. Li and KS~Zhang,
\newblock ``Optimal linear estimation fusion¡ªpart iv: Optimality and
  efficiency of distributed fusion,''
\newblock in {\em Proc. 2001 International Conf. on Information Fusion}, 2001.

\bibitem{chong1985information}
C.Y. Chong, S.~Mori, and K.C. Chang,
\newblock ``Information fusion in distributed sensor networks,''
\newblock in {\em American Control Conference, 1985}. IEEE, 1985, pp. 830--835.

\bibitem{chang1997optimal}
K.C. Chang, R.K. Saha, and Y.~Bar-Shalom,
\newblock ``On optimal track-to-track fusion,''
\newblock {\em Aerospace and Electronic Systems, IEEE Transactions on}, vol.
  33, no. 4, pp. 1271--1276, 1997.

\bibitem{drummond1997hybrid}
O.E. Drummond,
\newblock ``Hybrid sensor fusion algorithm architecture and tracklets,''
\newblock in {\em Proceedings of SPIE}, 1997, vol. 3163, p. 485.

\bibitem{willsky1982combining}
A.~Willsky, M.~Bello, D.~Castanon, B.~Levy, and G.~Verghese,
\newblock ``Combining and updating of local estimates and regional maps along
  sets of one-dimensional tracks,''
\newblock {\em Automatic Control, IEEE Transactions on}, vol. 27, no. 4, pp.
  799--813, 1982.

\bibitem{miller1998tracklets}
M.D. Miller, O.E. Drummond, and A.J. Perrella,
\newblock ``Tracklets and covariance truncation options for theater missile
  tracking,''
\newblock in {\em Proc. 1998 International Conf. on Multisource-Multisensor
  Data Fusion (FUSION¡¯98)}, 1998.

\bibitem{julier1997non}
S.J. Julier and J.K. Uhlmann,
\newblock ``A non-divergent estimation algorithm in the presence of unknown
  correlations,''
\newblock in {\em American Control Conference, 1997. Proceedings of the 1997}.
  IEEE, 1997, vol.~4, pp. 2369--2373.

\bibitem{niehsen2002information}
W.~Niehsen,
\newblock ``Information fusion based on fast covariance intersection
  filtering,''
\newblock in {\em Information Fusion, 2002. Proceedings of the Fifth
  International Conference on}. IEEE, 2002, vol.~2, pp. 901--904.

\bibitem{franken2005improved}
D.~Franken and A.~Hupper,
\newblock ``Improved fast covariance intersection for distributed data
  fusion,''
\newblock in {\em Information Fusion, 2005 8th International Conference on}.
  Ieee, 2005, vol.~1, pp. 7--pp.

\bibitem{hurley2002information}
Michael~B Hurley,
\newblock ``An information theoretic justification for covariance intersection
  and its generalization,''
\newblock in {\em Information Fusion, 2002. Proceedings of the Fifth
  International Conference on}. IEEE, 2002, vol.~1, pp. 505--511.

\bibitem{weng2012}
Z~Weng and P.M. Djuri\'{c},
\newblock ``A {Bayesian} approach to covariance estimation and data fusion,''
\newblock in {\em EUSIPCO}, 2012, vol.~1, p.~1.

\bibitem{james1964distributions}
A.T. James,
\newblock ``Distributions of matrix variates and latent roots derived from
  normal samples,''
\newblock {\em The Annals of Mathematical Statistics}, pp. 475--501, 1964.

\bibitem{anderson2003introduction}
T.W. Anderson,
\newblock {\em An introduction to multivariate statistical analysis.},
\newblock Wiley Series in Probability and Statistics, 2003.

\bibitem{gupta2000matrix}
A.K. Gupta and D.K. Nagar,
\newblock {\em Matrix variate distributions}, vol. 104,
\newblock Chapman \& Hall/CRC, 2000.

\end{thebibliography}

\end{document}